*Article*

# Probability Distribution Functions of Solar and Stellar Flares

**Takashi Sakurai** [1,*]

[1] National Astronomical Observatory of Japan, 2-21-1 Osawa, Mitaka, Tokyo 181-8588, Japan; sakurai.takashi@nao.ac.jp

\* Correspondence: sakurai.takashi@nao.ac.jp

**Abstract:** We studied the soft X-ray data of solar flares and found that the distribution functions of flare fluence are successfully modeled by tapered power law or gamma function distributions whose power exponent is slightly smaller than 2, indicating that the total energy of the flare populations is mostly contributed from a small number of large flares. The largest possible solar flares in 1000 years are predicted to be around X70 in terms of the GOES flare class. We also studied superflares (more energetic than solar flares) from solar-type stars, and found that their power exponent in the fitting of the gamma function distribution is around 1.05, much flatter than solar flares. The distribution function of stellar flare energy extrapolated downward does not connect to the distribution function of solar flare energy.

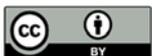

## 1. Introduction

Solar flares are the most energetic phenomena among a wide variety of magnetic activities taking place on the Sun [1, 2]. The first flare observation was made by Carrington in 1859 in white light [3], and the earlier observations were restricted to H$\alpha$ emission. Later radio and X-ray observations revealed that flares heat the corona from its normal 1-2 MK state to 10 MK or beyond. Then it was found in 1980s that the largest fraction of energy in flares goes to the kinetic energy of coronal mass ejections (CMEs). The most energetic flares liberate energies up to $10^{33}$ erg [4]. Now it is almost established that the energy release in solar flares is due to magnetic reconnection [5].

The number of flares $f(E)dE$ with energies between $E$ and $E+dE$ is distributed roughly in power law [6, 7] with a negative exponent, $f(E) \sim E^{-\alpha}$, namely smaller flares are more numerous. This property led Parker [8] to propose that the solar corona might be heated [9] by energies supplied from numerous small flares, later called nanoflares (in contrast to another class of theory based on waves [10, 11]). Here the important criterion is whether the power law index $\alpha$ is larger or smaller than 2 [12]. The total energy brought by all the flares with energies between $E_1$ and $E_2$ is

$$W = \int_{E_1}^{E_2} E\, f(E)\, dE \sim \frac{1}{2-\alpha}(E_2^{2-\alpha} - E_1^{2-\alpha}). \tag{1}$$

If $\alpha > 2$ then contributions from smaller events ($E_1 \to 0$) determine the total energy involved in the flare phenomenon. However, for the observed flares of moderate or large sizes it is generally believed that $\alpha < 2$ (1.8 or so). Therefore, more contributions to $W$ come from larger flares. On the large energy end, in terms of space weather it has been discussed how frequent very large (extreme) events would be [4]. Recent discovery of energetic flares (superflares) from solar-type (old and slowly rotating) stars [13, 14] has stimulated interest in solar extreme events.

In this article we will look into the probability distribution functions of solar X-ray flares. As numerical measures we use both X-ray peak flux and X-ray fluence which is the time integration of X-ray flux. The peak X-ray flux may be subject to detector saturation for very large flares (X17 or larger [15]), but we expect that X-ray fluence data would suffer less influences because of time integration. The data we use are the soft X-ray observation by the GOES satellites [16] (Figure 1). In Figure 1 are plotted $7.9\times10^4$ flare events of peak flux above $10^{-7}$ W m$^{-2}$ (1975–2020) and $3.4\times10^4$ events of fluence above $5\times10^{-5}$ J m$^{-2}$ (1997–2020).

In a foregoing study [17], the power law indices of $\alpha=2.11$ for X-ray peak flux and $\alpha=2.03$ for X-ray fluence were derived. Also they pointed out that $\alpha$ would be 1.88 if the background intensity is subtracted from the flare data.

Our conclusions are that, although the distributions are roughly power law, at the high energy end they are tapered off exponentially. The power law is a scale-free distribution, and could be a viable model for phenomena taking place in scales sufficiently smaller than the entire system (the Sun in the present case). Solar flares are powered by magnetic energy stored in sunspot regions, and since sunspot sizes cannot exceed the Sun (or more effectively limited by the depth of the convection zone), a limit is expected on the amount of energy released in flares. Therefore, a power-law extension of the distribution overestimates the frequency of extremely large solar flares. We will briefly touch upon the energy distribution function of stellar flares and compare it with the solar flare case in Section 4.

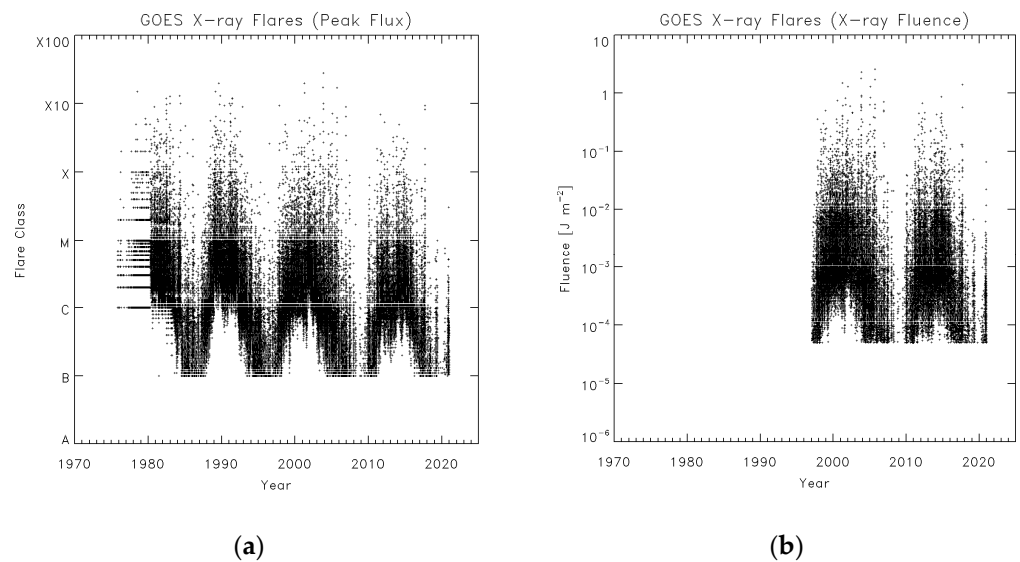

(**a**) (**b**)

**Figure 1.** Solar soft X-ray flares detected by the GOES satellites. Panels (a) and (b) show, respectively, peak flux and fluence of flares as a function of year.

## 2. Materials and Methods

We use the X-ray flux observed in the 0.1-0.8 nm band of the GOES X-ray sensors (1980–2020). The peak flux values are traditionally expressed [18] in terms of A ($10^{-8}$ W m$^{-2}$), B ($10^{-7}$ W m$^{-2}$), C ($10^{-6}$ W m$^{-2}$), M ($10^{-5}$ W m$^{-2}$), and X ($10^{-4}$ W m$^{-2}$) classes. Namely, an M5.5 flare means a flare of peak flux $5.5\times10^{-5}$ W m$^{-2}$, and an X12 flare has a peak flux of $12\times10^{-4}$ W m$^{-2}$. The data are available from 1975, but the data before 1980 had one-digit accuracy (C2, M4, etc.) so that we use the data after 1980 which have two-digit accuracy (C2.1, M3.9, etc.). The data after 1997 also include the fluence values (time integration of X-ray flux). In the following analysis we have picked up flares with peak flux above M3 or $3\times10^{-5}$ W m$^{-2}$ (1980-2020, 1720 events) and also flares with fluence above $1\times10^{-2}$ J m$^{-2}$ (1997-2020, 1945 events). We have used the old operational scale instead of the new science scale adopted by NOAA [15] (operational values = 0.7 × science values). The peak flux data of 2020 and the background data of 2011–2020 which were in the science database had been converted to the operational scale.

As previously suggested [17], background subtraction may affect significantly the data from small events particularly. Here we have adopted a simplified approach and subtracted the daily background value from the flare peak flux, and the value of the background times the duration from the fluence data (the duration of a flare can be computed from the flare start and end times given in the database). The background subtraction was not applied for 1980-1983 April (data not available) and on other days with no background data available, and when the background exceeds the flare intensity (e.g. when the background was too high because of a preceding flare).

We will work on the complementary cumulative distribution function CCDF($F$) of fluence $F$ defined as [19]

$$\mathrm{CCDF}(F) = n(\mathrm{fluence} \geq F)/N, \qquad (2)$$

where $N=1945$ is the total number of events under study. The probability distribution function (PDF) is defined by

$$\mathrm{PDF}(F) = -\frac{\mathrm{d(CCDF)}}{\mathrm{d}F}, \qquad (3)$$

but here we will use the flare occurrence rate defined by

$$f(F) = \mathrm{PDF}(F)\,\frac{N}{\tau_{\mathrm{obs}}}, \qquad (4)$$

where $\tau_{\mathrm{obs}}$ is the total duration of data (24 years). The dimension of $f(F)$ is 1/(J m$^{-2}$ day). Figure 2 shows CCDF($F$) (a) and $f(F)$ (b).

We can likewise define the quantities for X-ray peak flux $F_{\mathrm{P}}$, with $N_{\mathrm{P}}=1720$ and $\tau_{\mathrm{obs,p}}=41$ years. The dimension of $f(F_{\mathrm{P}})$ is 1/(W m$^{-2}$ day).

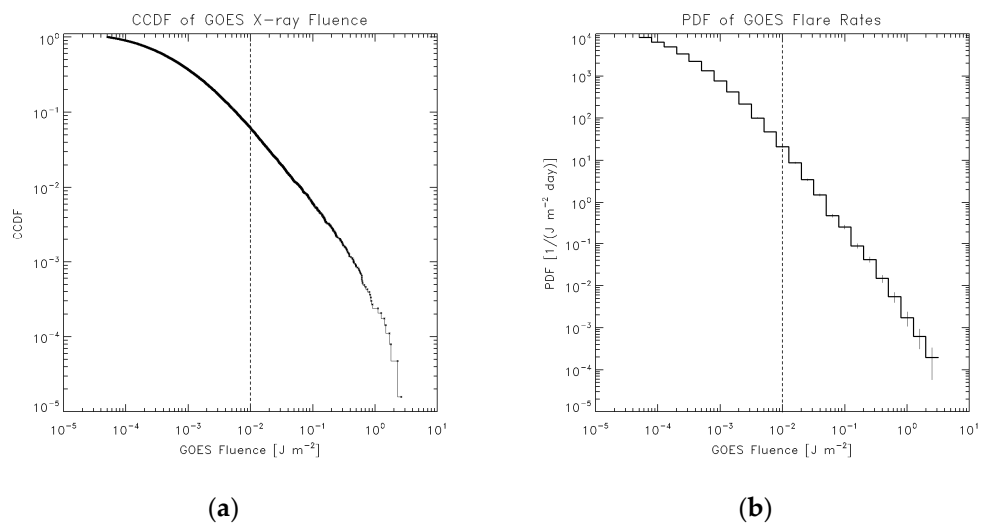

(a) (b)

**Figure 2.** Complementary cumulative distribution function (**a**) and occurrence rates (**b**) of flare fluence $F$. The histogram in panel (**b**) is made with a bin size of $\Delta\log_{10}F=0.2$. Short vertical bars indicate statistical errors of sqrt($n$) on the bins with counts $n$.

We try to fit the observed CCDF by four kinds of models; power law, tapered power law, gamma function, and Weibull distributions. The power law distribution is defined by [19]

$$\text{CCDF}(F) = \left(\frac{F}{F_0}\right)^{-\alpha+1}, \quad \text{PDF}(F) = \frac{\alpha-1}{F_0}\left(\frac{F}{F_0}\right)^{-\alpha}, \tag{5}$$

where $\alpha > 1$ and $F_0$ (=$1\times10^{-2}$ J m$^{-2}$; $F_{0p}=3\times10^{-5}$ W m$^{-2}$ for the peak flux data) is the lower boundary of the fitting. The other three are two-parameter models. The tapered power law distribution is defined by [20]

$$\text{CCDF}(F) = \left(\frac{F}{F_0}\right)^{-\alpha+1} \exp\left[-\beta \frac{F-F_0}{F_0}\right], \tag{6}$$

where $\alpha > 1$, $\beta > 0$. The gamma function distribution is defined by [20]

$$\text{PDF}(F) = \frac{C}{F_0}\left(\frac{F}{F_0}\right)^{-\alpha} \exp\left[-\beta \frac{F-F_0}{F_0}\right], \quad C = \frac{\beta^{1-\alpha}}{F_0 \Gamma(1-\alpha, \beta)}, \tag{7}$$

where $\alpha > 1$, $\beta > 0$, and $\Gamma$ is the incomplete gamma function. The CCDF of the gamma function distribution is given as

$$\text{CCDF}(F) = \frac{\Gamma(1-\alpha, \beta F/F_0)}{\Gamma(1-\alpha, \beta)}. \tag{6}$$

The tapered power law and gamma function distributions have been used in geophysics in representing the distributions of earthquake magnitudes [21], in contrast to the power law distribution which is called the Gutenberg-Richter relation in seismology [22, 23].

The Weibull distribution is defined by

$$\text{CCDF}(F) = \exp\left[-\beta \left(\frac{F}{F_0}\right)^k + \beta\right], \tag{8}$$

where $k > 0$, $\beta > 0$. The Weibull distribution was first introduced to evaluate the failure rates of industrial products [24].

The parameters can be determined by using the maximum likelihood method, namely by maximizing the log-likelihood (LLH) defined by

$$\text{LLH} = \sum_{i=1}^{N} \ln \text{PDF}(F_i). \tag{9}$$

Specific forms of the maximum-likelihood solutions can be found in the literature for power law [19], tapered power law [25], gamma function [20], and Weibull [26] distributions. The goodness of fit can be evaluated by the Kolmogorov-Smirnov (K-S) test [27], which uses the maximum difference between the observed and theoretical CCDFs. Whether one model is superior to the others can be estimated by Akaike's Information Criterion (AIC) [28] given by

$$\text{AIC} = -2 \cdot \text{LLH} + 2K, \tag{10}$$

where $K$ is the number of parameters used for fitting ($K=1$ for power law, $K=2$ for the other three models). By introducing more parameters, the fitting will improve and one may get larger LLH, but meaningful improvement requires that AIC decreases sufficiently (reduction in AIC of about 9-11 [29]). A similar analysis was conducted by this author on the area distribution of sunspots [30].

Statistical errors in the determined parameter values can be estimated by generating many statistical distributions from the same parameter values and by seeing the distributions of the fitted parameter values (parametric bootstrap method [31]). The errors given in Tables 1 and 2 are the one-sigma errors thus derived.

## 3. Results

Tables 1 and 2 summarize the fitting results for X-ray fluence and peak flux, respectively. ΔAIC is the difference in AIC from the smallest AIC value (indicating the most likely model) among all the models, and models with ΔAIC ≳ 9–11 are poorly supported compared to the ΔAIC=0 model. Therefore, we can conclude that the power law model, shown in Figure 3, is not favorable compared to the other two-parameter models.

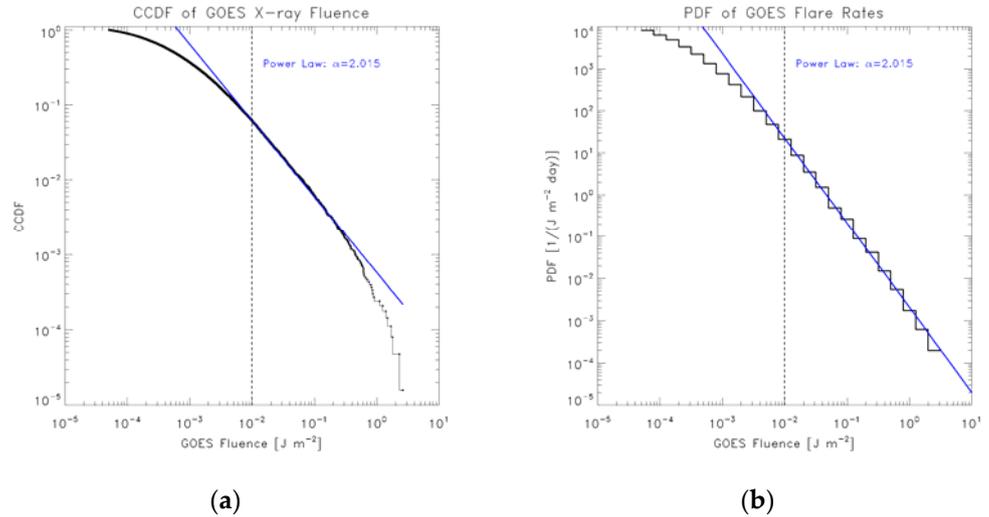

(**a**) (**b**)

**Figure 3.** (**a**) The blue line shows the power-law fit to the observed CCDF of GOES flare fluence data. (**b**) The derived power-law distribution overplotted on the observed flare occurrence rate histogram. In both panels the vertical dashed line indicates the lower boundary (1×10$^{-2}$ J m$^{-2}$) of data used for fitting.

A model can be rejected by the K-S test by examining its measure, $\sqrt{N}$ × KS, where KS is the maximum difference between the observed and model CCDFs. The probability that the observed $\sqrt{N}$ ×KS value or larger is obtained (the "KS P-value" in Tables 1 and 2) is given analytically as the Kolmogorov-Smirnov function which has only a weak dependence on $\sqrt{N}$ regardless of the models assumed [32]. If the value of $\sqrt{N}$ × KS is large and the corresponding KS P-value is small, we can conclude that the model is rejected. However, we found that the KS measures are generally small and the P-values are not so small even for the power-law models. This happens because the K-S test is not sensitive to misfitting at the tail of the distributions. If we reduce $F_0$, the fitting degrades and eventually all the models tend to be rejected by the K-S test. The ambiguities in setting the lower bound $F_0$ of the probability distribution function will be discussed in Section 4.

### 3.1. Tapered Power Law and Gamma Function Distributions

These two models (Figures 4(a), (b)) show similar performance in terms of ΔAIC. The power-law indices $\alpha$ for the fluence are 1.973 for the tapered power law model and 1.949 for the gamma function model. Considering one-sigma errors we may still say that the values of $\alpha$ are less than 2, namely the total energy of all the flare populations is contributed primarily from large events. The values of $\alpha$ for the peak flux are 2.077 for the tapered power law model and 2.040 for the gamma function model, larger than 2. These values generally agree with [17].

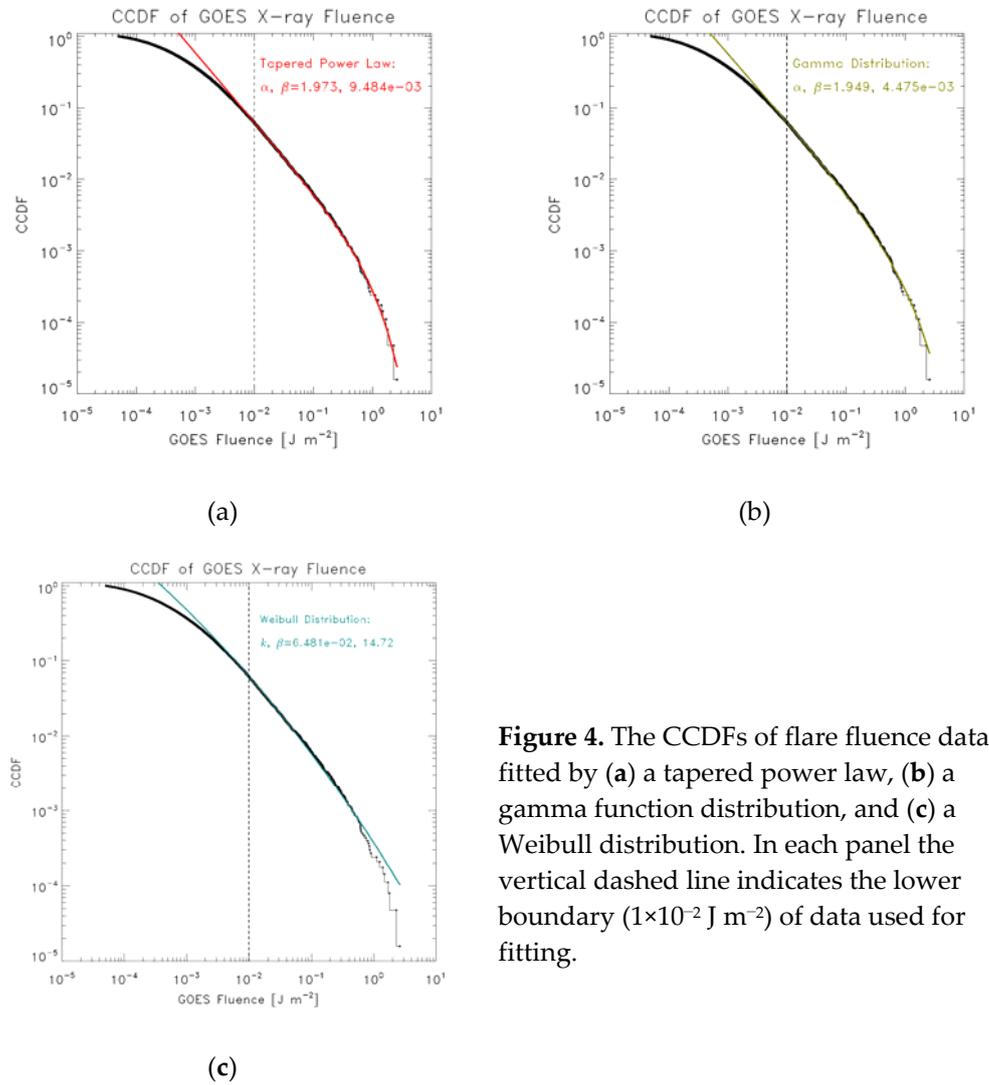

**Figure 4.** The CCDFs of flare fluence data fitted by (**a**) a tapered power law, (**b**) a gamma function distribution, and (**c**) a Weibull distribution. In each panel the vertical dashed line indicates the lower boundary ($1\times10^{-2}$ J m$^{-2}$) of data used for fitting.

### 3.2. Weibull Distribution

In terms of ΔAIC, the Weibull distribution (Figure 4(c)) cannot be dismissed but is supported only marginally (30 times less likely than the ΔAIC=0 model).

If we extend the distribution to $F \ll F_0$, the Weibull PDF approaches a power law with exponent $1-k$. Therefore, the fluence PDF will behave like $F^{-(1-k)}=F^{-0.935}$, which is significantly flatter than the tapered power law and gamma function distributions. This is another reason why the Weibull distribution is not favored compared to the tapered power law and gamma function distributions.

**Table 1.** Derived parameters for X-ray fluence distributions ($F_0=1\times10^{-2}$ J m$^{-2}$, $N=1945$).

| Model | α | β | ΔAIC | $\sqrt{N}\times$KS | K-S P-value |
|---|---|---|---|---|---|
| Power law | 2.015±0.023 | | 11.6 | 0.69 | 0.72 |
| Tapered power law | 1.973±0.021 | 0.00948±0.0031 | 0.00 | 0.47 | 0.98 |
| Gamma function | 1.949±0.032 | 0.00448±0.0020 | 1.10 | 0.57 | 0.90 |
| Weibull | $k$=0.0648±0.0242 | 14.7±5.9 | 7.20 | 0.69 | 0.73 |

**Table 2.** Derived parameters for X-ray peak flux distributions ($F_{0P}=3\times10^{-5}$ W m$^{-2}$, $N=1720$).

| Model | α | β | ΔAIC | √N ×KS | K-S P-value |
|---|---|---|---|---|---|
| Power law | 2.162±0.028 | | 15.7 | 0.89 | 0.41 |
| Tapered power law | 2.077±0.032 | 0.026±0.007 | 0.00 | 0.40 | 0.99 |
| Gamma function | 2.040±0.045 | 0.014±0.005 | 1.48 | 0.44 | 0.99 |
| Weibull | $k$=0.0104±0.030 | 10.2±3.3 | 7.00 | 0.57 | 0.90 |

*3.3. Comparison with Published Results*

Similar analyses have been made by Veronig et al. [17] using the power law model (Appendix A) and by Gopalswamy [33] using the Weibull function model (Appendix B). Figure 5 compares the present results with these publications. Our tapered power law and gamma function distributions show similar behaviors. The power law distributions derived by [17] roughly follow our data histograms but deviates from the data at the higher end of the data. The Weibull distributions (both ours and of [33]) become flatter at the smaller ends and decay less at the higher ends.

In the following we will only consider the gamma function distributions. The Weibull distributions are less favorable as described before. The tapered power law PDFs are made of a mixture of two power-law exponents $\alpha$ and $\alpha -1$, and this sometimes may lead to undesirable features [30].

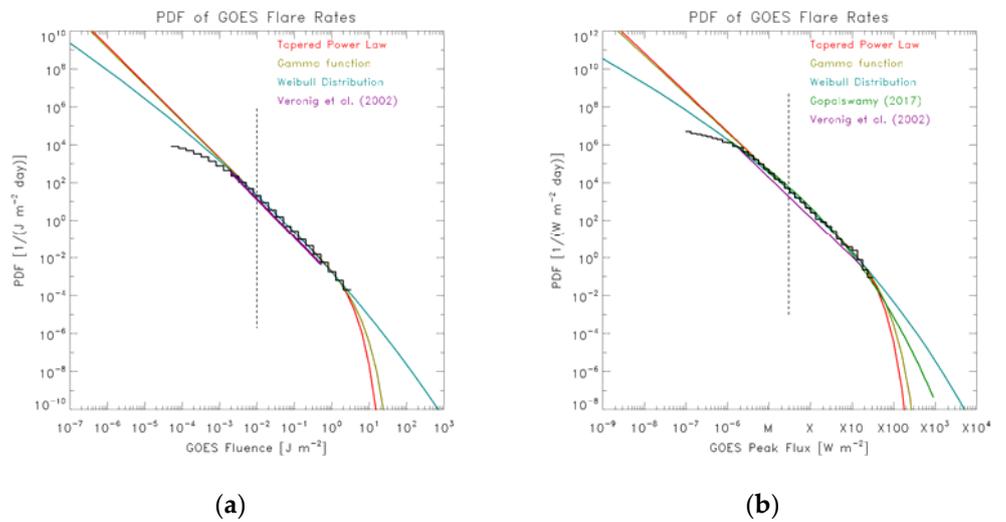

(a) (b)

**Figure 5.** Our analysis results are compared with previous publications. Panels (a) and (b), respectively, show the flare occurrence rates (PDFs) for fluence and peak flux. The thick black histograms are the observed PDFs. Red, olive, and teal curves indicate tapered power law, gamma function, and Weibull distributions fitted to the data. Purple and green curves are power law [17] and Weibull [33] fits in the previous publications. The vertical dashed lines indicate the lower boundary in data used for fitting, namely $1\times10^{-2}$ J m$^{-2}$ for fluence and M3 ($3\times10^{-5}$ W m$^{-2}$) for peak flux, respectively.

*3.4. Prediction of Extreme Events*

Based on our fit to the flare fluence data in terms of the gamma function distribution, we will estimate the expected rates of large flares. The total soft X-ray energy emitted, $E_X$, is given in terms of the X-ray flare fluence $F$ as

$$E_X = 2\pi \times (1 \text{ au})^2 \times F \times 10^7. \tag{11}$$

Here $F$ is in J m$^{-2}$, $E_X$ is in erg, and 1 astronomical unit (au) = 1.5×10$^{11}$ m. From this we may estimate $E_{rad}$, the total radiated energy all over the electromagnetic spectrum (contributed mostly from UV) in erg, as [34]

$$E_{rad} = 1.03 \times 10^9 \times E_X^{0.766}, \tag{12}$$

and roughly we assume that the flare total energy (kinetic plus radiative) is 4 times this quantity [34, 35]. Finally, the quantities derived from the flare fluence data can be translated into the flare peak flux $F_P$ (J m$^{-2}$) roughly by

$$F_P \approx 7.93 \times 10^{-4} \times F^{0.945}, \tag{13}$$

which is the relation derived from our data.

Using these relations, we have translated our probability distribution function for soft X-ray fluence to flare peak flux and total flare energy as in Tables 3a and 3b. Table 3a gives, for a specified value of flare fluence, the approximate peak flux, energy, and interval of flares exceeding the specified fluence. Table 3b gives, for a specified interval of a flare, the flare fluence, peak flux, and energy. The numbers are based on the gamma function distribution of $\alpha$=1.949, $\beta$=0.00448. The second numbers after a slash are derived by assuming one-sigma errors ($\alpha$=1.949+0.021, $\beta$=0.00448–0.0010 because the errors are correlated).

**Table 3a.** Predicted flare intervals as a function of X-ray fluence values (gamma function distribution, $\alpha$=1.949±0.032, $\beta$=0.00448±0.0020). The second numbers after a slash are derived by assuming one-sigma errors ($\alpha$=1.949+0.032, $\beta$=0.00448-0.0020).

| X-ray fluence (J m$^{-2}$) | Approx. GOES flux | Total energy (erg) | Interval (years) |
|---|---|---|---|
| 1.0×10$^{-2}$ | M1.0 | 1.5×10$^{31}$ | 1.3×10$^{-2}$/1.3×10$^{-2}$ |
| 1.0×10$^{-1}$ | M9.0 | 8.8×10$^{31}$ | 1.4×10$^{-1}$/1.4×10$^{-1}$ |
| 1.0×10$^{0}$ | X7 | 5.1×10$^{32}$ | 3.0×10$^{0}$/2.3×10$^{0}$ |
| 2.0×10$^{0}$ | X15 | 8.7×10$^{32}$ | 1.2×10$^{1}$/7.3×10$^{0}$ |
| 5.0×10$^{0}$ | X36 | 1.8×10$^{33}$ | 1.8×10$^{2}$/5.6×10$^{1}$ |
| 6.3×10$^{0}$ | X45$^\dagger$ | 2.1×10$^{33}$ | 4.6×10$^{2}$/1.1×10$^{2}$ |

$^\dagger$ Carrington event (1859 Sept.1) [34].

**Table 3b.** Predicted flare sizes as a function of flare intervals (gamma function distribution, $\alpha$=1.949±0.032, $\beta$=0.00448±0.0020). The second numbers after a slash are derived by assuming one-sigma errors ($\alpha$=1.949+0.032, $\beta$=0.00448-0.0020).

| X-ray fluence (J m$^{-2}$) | Approx. GOES flux | Total energy (erg) | Interval (years) |
|---|---|---|---|
| 6.5×10$^{0}$/9.9×10$^{0}$ | X46/X69 | 2.2×10$^{33}$/3.0×10$^{33}$ | 1.0×10$^{3}$ |
| 1.0×10$^{1}$/1.6×10$^{1}$ | X70/X108 | 3.0×10$^{33}$/4.3×10$^{33}$ | 1.0×10$^{4}$ |
| 1.4×10$^{1}$/2.3×10$^{1}$ | X95/X152 | 3.8×10$^{33}$/5.6×10$^{33}$ | 1.0×10$^{5}$ |
| 1.8×10$^{1}$/3.0×10$^{1}$ | X122/X197 | 4.7×10$^{33}$/6.9×10$^{33}$ | 1.0×10$^{6}$ |
| 2.7×10$^{1}$/4.6×10$^{1}$ | X177/X293 | 6.3×10$^{33}$/9.5×10$^{33}$ | 1.0×10$^{8}$ |
| 3.4×10$^{1}$/5.9×10$^{1}$ | X224/X374 | 7.7×10$^{33}$/1.2×10$^{34}$ | 4.6×10$^{9}$ |

From this table we find that the Carrington event in 1859 September 1, which is estimated to be an X45 flare [34], may take place every 460 years, or every 110 years if we allow one-sigma errors in $\alpha$ and $\beta$. The largest flare in 1000 years would be X46, or X69 (or X70 as a round number) if we allow one-sigma errors. If we wait for the age of the solar system (4.6×10$^9$ years), the largest flare we experience would be X224, or X374 if we allow one-sigma errors.

## 4. Discussion

Recent discovery of energetic flares (called superflares) on solar-type stars observed with the Kepler satellite [36] (whose main targets are exoplanets, by high-precision photometry to detect the brightness decrease due to the transit of planets) has attracted attention in relation to extreme events that might happen on our Sun [13, 14, 37, 38]. Those stellar flares are observed in the visible light; in the case of solar flares the emission in the visible light is detected only from intense flares (called "white-light flares") [1].

As an application of the analysis methods described in this paper, we have studied the energy distribution functions of stellar flares. By using the most recent database on superflares from solar-type stars (G-type main-sequence stars with effective temperature of 5600−6000 K and rotation period longer than 20 days) which carefully removed contaminations from binaries or evolved stars [39], we obtained Figure 6 (panel (a) for CCDF and panel(b) for PDF). The data contain 2341 flares from 265 stars observed over 4 years. Both power law (blue) and gamma function (olive) model fits are shown. Here the assumption is that the observed distribution function may mimic a hypothetical distribution function of flares from a single Sun-like star observed over 4×265=1060 years.

The value of ΔAIC of the power law model is about 90 compared to the gamma function distribution, so that the power law is unfavorable compared to the gamma function distribution. The power law model gives a K-S measure √N × KS = 4.2 and its P-value is infinitesimally small. The gamma function model gives the parameter values $\alpha$=1.05, $\beta$=0.16. The P-value from the K-S test is 0.39 and not so high, indicating room for better model of distribution functions.

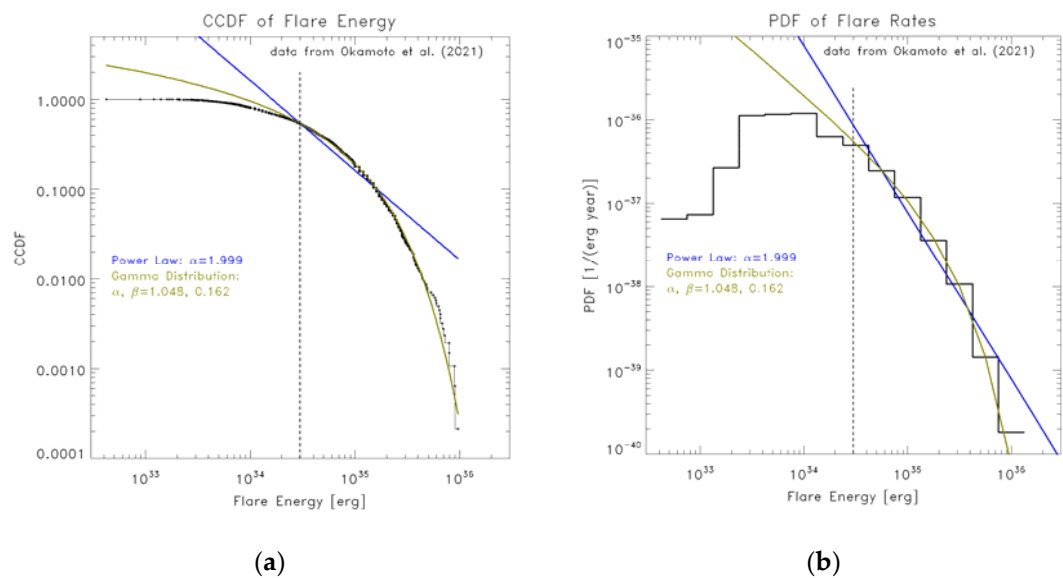

(**a**) (**b**)

**Figure 6.** Statistical distributions of flares on solar-type stars. Panel (**a**) shows CCDF and a power-law (blue) and gamma function distribution (olive) fits, respectively. Panel (**b**) shows the histogram of flare occurrence rates and the derived power law and gamma function distributions. The data are from [38]. In each panel the vertical dashed line indicates the lower boundary (3×10$^{34}$ erg) of data used for fitting.

If we combine Figure 6 and Figure 4b after converting flare fluence to radiated energy in the latter, we obtain Figure 7. The $F_0$ value of the GOES flare fluence, $F_0$=1×10$^{-2}$ J m$^{-2}$, is translated to the radiated energy of 3.8×10$^{30}$ erg. It can be clearly seen that the PDF of solar flares does not connect to the stellar flare PDF but decays at an energy of approximately $10^{33.5}$ erg. Likewise, the PDF of stellar flares has a flatter distribution and does not connect to the PDF of solar flares. Although solar and stellar flares are both believed to be powered by magnetic energy (most likely by magnetic reconnection) of spotted regions, the distributions of the size and magnetic field strength in starspots would not be the same as in sunspots and may have a wider variety, because of different internal structures and rotation periods. Stellar flares may have a variety of power-exponent values, and the present results of $\alpha$=1.95 for solar flares and $\alpha$=1.05 for stellar flares are just two examples and there might be a continuous distribution of $\alpha$ between 1 and 2.

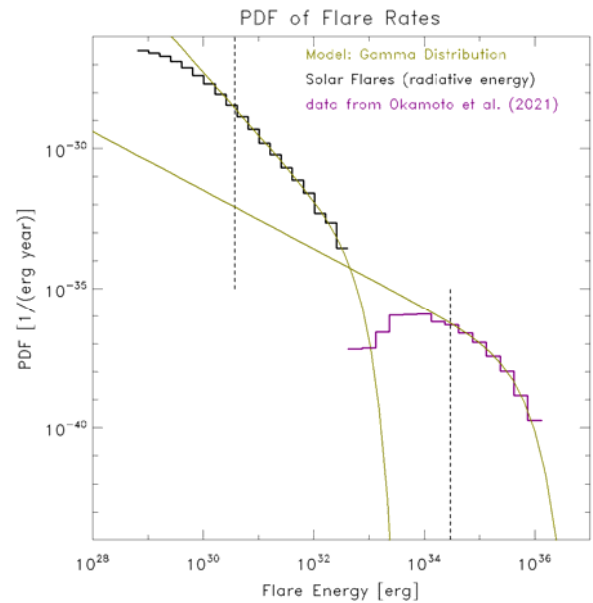

**Figure 7.** Comparison of solar and stellar flare energy distributions. Black and purple histograms show the occurrence rates of solar X-ray flares and stellar flares [38]. The olive curves are fit to these data by the gamma function distributions. The vertical dashed lines indicate the lower boundary for fitting adopted for the solar flare data ($3.8\times10^{30}$ erg of radiated energy) and for the stellar flare data ($3\times10^{34}$ erg).

These results depend on the assumed values of the lower boundary of data for fitting. Ref. [19] proposes that the value of the lower boundary may also be selected by minimizing the K-S metric, but it seems that no consensus has been obtained yet [39, 40]. In our studied cases, the turning down of the distributions toward smaller flares is due to the detection limits; small solar flares are not recognized as flares when the background level rises in the activity maximum period of the Sun's eleven-year activity cycle. In the case of stellar flares, weak flares in more distant stars may be missed. On the other hand, if we set the boundary to a very large value, we will be left with only a small number of samples. Our present approach is to find a compromise so that the samples are well above the detection limit and still a large number of samples is retained. It is desirable that more subjective methods are developed. It may also be useful to analyze data on flares from individual stars rather than using a composite data made of many stars as we did in this study.

## 5. Conclusions

From the analysis of soft X-ray data of the GOES satellites, we found that the distribution functions of solar flare fluence are successfully modeled by tapered power law or gamma function distributions whose power exponent is slightly smaller than 2, indicating that the total energy of the flare populations is mostly contributed from a small number of large flares. The tapering off of the distribution is statistically significant and the power law fit is rejected. The largest possible flares in 1000 years are predicted to be around X46, or X69 if we allow one-sigma errors in the derived parameters. The upper limits of X224 (X374 if one allows one-sigma errors) are derived even by considering the lifetime of the solar system.

Similar treatment was applied to flares from solar-type stars. They were fitted by a gamma function distribution (the power law is rejected), but their power exponent is around 1.05, smaller than the solar flare case (around 1.95). Therefore, the stellar flare data analyzed here do not connect to the energy distribution function of solar flares. It would be interesting to see how the distribution functions differ among solar-type stars with different effective temperatures and rotation periods.

**Supplementary Materials:** The data sets we used in the present analysis are available at

http://solarwww.mtk.nao.ac.jp/sakurai/tar/goesfluence_datalist.zip

http://solarwww.mtk.nao.ac.jp/sakurai/tar/goesflux_datalist.zip

for the flare fluence above $5\times10^{-5}$ J m$^{-2}$ (1997-2020) and peak flux above $1\times10^{-7}$ W m$^{-2}$ (1980-2020), respectively.

**Funding:** This research was supported by JSPS KAKENHI Grant numbers JP15H05816 and JP20K04033.

**Data Availability Statement:** The GOES flare peak flux data are available at

https://www.ngdc.noaa.gov/stp/space-weather/solar-data/solar-features/solar-flares/x-rays/goes/xrs/
for the years 1975-2016, and at

tftp://ftp.swpc.noaa.gov/pub/warehouse/

for 2017 and later. The fluence data are included in these data after 1997. The X-ray background data for the years 1983-2011 are given in

ftp://ftp.ngdc.noaa.gov/STP/SOLAR_DATA/SATELLITE_ENVIRONMENT/Daily_Fluences/Fluence.

The background data for 2011-2019 are available at

https://satdat.ngdc.noaa.gov/sem/goes/data/science/xrs/

and at

https://data.ngdc.noaa.gov/platforms/solar-space-observing-satellites/goes/

for 2020. The background values between 1983 March and 1987 were calculated from the 1-minute intensities given at

https://satdat.ngdc.noaa.gov/sem/goes/data/avg.

The data on stellar flares discussed in Section 4 are provided by [38] at the following URL.

https://cfn-live-content-bucket-iop-org.s3.amazonaws.com/journals/0004-637X/906/2/72/revision1/apjabc8f5t2_mrt.txt?AWSAccessKeyId=AKIAYDKQL6LTV7YY2HIK&Expires=1663922680&Signature=f9%2B%2FXGkDmwCeg2WIQr6XrGm8l80%3D

**Acknowledgments:** The author would like to thank Shin Toriumi for discussion. It is a great pleasure for the author to express appreciations to Marcel Goossens for his hospitality during the author's visit to KU Leuven in 1989 that led to a couple of jointly-written papers.

**Conflicts of Interest:** The author declares no conflict of interest.

**Appendix A**

Veronig et al.'s [17] parameters of power-law fits shown in Figure 5 are, for flare fluence; $N$=8400, $\tau_{obs}$=4 years, $\alpha$=2.03, and $F_0$=2x10$^{-3}$ J m$^{-2}$. For flare peak flux, $N_p$=49409, $\tau_{obs,p}$=25 years, $F_{0p}$=2x10$^{-6}$ W m$^{-2}$, and $\alpha$=2.11.

**Appendix B**

Gopalswamy's [33] parameters of Weibull distribution fit shown in Figure 5 are; $N_p$=55285, $\tau_{obs,p}$=48 years, $F_{0p}$=9.1x10$^{-7}$ W m$^{-2}$, $k$=0.167, and $\beta$=3.77.